\documentclass[epjST]{svjour}
\usepackage{graphics}
\usepackage{graphicx}
\usepackage{cite}
\begin{document}
\def\beq{\begin{equation}}
\def\endeq{\end{equation}}
\def\begdi{\begin{displaymath}}
\def\enddi{\end{displaymath}}
\def\ep{\varepsilon}
\def\speq{\hspace{1mm} = \hspace{1mm}}    
\def\hilight{\textbf}
\def\Z{\mathbb{Z}}
\def\half{\frac{1}{2}}
\def\defeq{\;\buildrel\hbox{\small def}\over{\,=}\;}    

\title{Tidal friction in satellites and planets. \protect\\ The new version of the creep tide theory}
\author{S.Ferraz-Mello \inst{1}\fnmsep\thanks{\email{sylvio@iag.usp.br}}
\and C.Beaug\'e \inst{2} \and H.A.Folonier \inst{1} \and G.O.Gomes \inst{1} }
\institute{Instituto de Astronomia, Geof\'{\i}sica e Ciências Atmosféricas, Universidade de S\~ao Paulo, Brasil \and
Observatorio Astron\'omico, Universidad Nacional de C\'ordoba, Argentina}
\abstract{
Paper on the creep tide theory and its applications to satellites and planets with emphasis on a new set of differential equations allowing easier numerical studies. The creep tide theory is a new paradigm that does not fix a priori the tidal deformation of the body, but considers the deformation as a low-Reynolds-number flow. The evolution under tidal forces is ruled by an approximate solution of the Navier-Stokes equation depending on the body's viscosity with no ad hoc assumptions on its shape and orientation. It reproduces closely the results of Darwinian theories in the case of gaseous planets and stars, but the results are completely different in the case of stiff satellites and planets. It explains the tidal dissipations of Enceladus and Mimas. The extension of the theory to nonhomogeneous icy satellites with a subsurface ocean allows the amplitude of the forced oscillations around synchronization (librations) to be better determined.}
 
\maketitle
\large
\section{Introduction}
\label{intro}
\begin{figure}[t]
	\begin{center}
		\resizebox{0.5\columnwidth}{!}{%
			\includegraphics{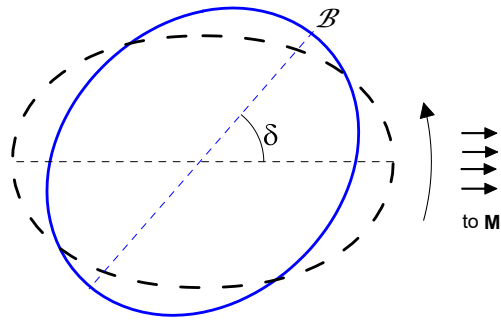} }
		\caption{Dynamic tide (solid blue) delayed w.r.t. the static tide (dashed black). The bulge ${\cal B}$ is dragged by the relative rotation of the body w.r.t. the attraction of the external body $\tens{M}$ }
		\label{fig:delay}
	\end{center}
\end{figure}

This paper deals with the theory proposed by Ferraz-Mello in 2012-2013 \cite {RhDDA, RhEPS, Rh1} for the study of the tidal evolution of planetary systems in the coplanar case, as an alternative to the classical Darwin theories. Under the action of an external gravitational field, one celestial body is deformed by tidal forces. If the body responds as a perfect fluid, two opposite bulges will form along the direction of the tidal forces and the shape of the body will be approximated by an ellipsoid (the so-called static tide). However, if the body is not a perfect fluid, it will offer some resistance to the deformation and, as a consequence, the deformation will be delayed and, if the orientation of the tidal forces changes with respect to the body, the deformation will not be fully deployed (see fig. \ref{fig:delay} ).
The consequences are twofold: the bulges will not have the same height as in the static tide and the resulting figure will not be oriented along the direction of the tidal forces but will show a deviation in the direction of the motion of the body with respect to the external force field.

In the classical versions of the Darwin theory, the deviation in the direction ($\delta$)
is small and the height of the bulges is multiplied by a response factor $\cos\delta$. This is sometimes called \textit{weak tide approximation} \cite {Ale}. Since $\delta$ is assumed to be a small quantity, the response factor $\cos\delta$ differs from 1 by a second-order infinitesimal, and this difference is omitted in almost all modern versions of Darwin theory. This approximation may lead to some interpretation difficulties in extreme cases (\cite{EfW09} Sec.9.1).  

In Darwinian theories, $\delta$ is an arbitrarily fixed small quantity. Strictly speaking, these theories introduce a time lag instead of a phase lag, and each term in the trigonometric series describing the shape of the body is delayed by an amount proportional to its frequency. This is the standard rheology of classical Darwin's theory (a.k.a. CTL or constant time lag theories)\footnote{For a synthetic account of the classical Darwin theory, see \cite{FRH}. In \cite{FRH}, Darwin theory is developed without specifying a particular law relating lags and frequencies and the different lags are just identified with different subscripts. The given generic equations are valid for various rheologies which may be introduced by just adopting one \textit{ad hoc} law relating lags and frequencies. As an example, in \cite{FRH}, Sec. 15, the equations are used to obtain the results for the particular case of linear theories with a constant time lag (CTL theories).}.  
In modern versions of Darwin's theories, different \textit{ad hoc} rheologies were used. For example, Efroimsky and Lainey \cite{EfLainey} have adopted a rheology based on the behavior of Earth's seismic waves in which phase lags and frequencies are related by an inverse power law. Inverse power laws bring an additional problem as the phase lag may become too large when the frequency tends to zero (as in spin-orbit synchronous motions) forcing the use of composite models in which the extended mass behaves as a Maxwell body at low frequencies \cite{MakE, Noy} . More recently, more complex rheologies arising from laboratory measurements and involving several free parameters, as the rheology associated with Andrade bodies, have been used to study the tidal evolution of stiff bodies \cite{Ef12a, Ef12b, Ef15} with good results.   
     
\section{The creep tide. Homogeneous bodies}\label{sec:creep}
The main difference between the creep tide theory and the classical Darwin's theory is that, in the creep tide theory, the lag $\delta$ and the response factor are not arbitrary parameters, but physical quantities determined from the first principles of Physics governing the motion of a Newtonian fluid when the flow's Reynolds number is low. They are given by the Newton's creep law, in the form
\beq\label{eq:creep}
\dot{\zeta}=-\gamma(\zeta-\rho)  
\endeq
where $\zeta(\widehat{\varphi},\widehat{\theta},t)$ is the distance to the center of gravity of the body of the surface point whose colatitude and longitude are $\widehat{\varphi},\widehat{\theta}$, $t$ is the time and $\rho(\widehat{\varphi},\widehat{\theta},t)$ is the distance of the corresponding point on the ideal triaxial ellipsoid which would be the surface body would it be homogeneous and inviscid, that is, if the body was, at each instant, in hydrostatic equilibrium (static tide).
In the coplanar case, this ideal ellipsoid is the composite of the Jeans prolate ellipsoid defined by the  tidal forces and a Maclaurin spheroid defined by the rotation of the body. The equatorial prolateness and polar oblateness of the resulting ellipsoid are, respectively, 
\beq\label{eq:prola}
\epsilon_\rho=\frac{a-b}{R_e} \simeq  \frac{15}{4}\left(\frac{M}{m}\right)\left(\frac{R}{r}\right)^3,
\endeq
\beq\label{eq:obla}
\epsilon_z=1-\frac{c}{R_e} \simeq \half\ \epsilon_\rho + \frac{5R^3\Omega^2}{4mG}.
\endeq
where $a,b,c$ are the ellipsoid semiaxes, $R$ is the mean radius, $R_e=\sqrt{ab}$ is the mean equatorial radius, $G$ is the gravitation constant, $m$ is the mass of the primary body, $M$ is the mass of the companion body creating the tidal forces that are acting on the primary, $\Omega$ is the angular rotation of the primary, $r$ is the distance from the primary to the companion.
The equation of the ideal triaxial ellipsoid is
\beq\label{eq:rho}
\rho=R_e \big(1 + \half \epsilon_\rho \sin^2\widehat\theta \cos (2\widehat\varphi - 2\varphi) - \epsilon_z \cos^2\widehat\theta \big),
\endeq
where $\varphi$ is the true longitude of the companion in its equatorial orbit around the primary\footnote{In the explicit expressions of $\epsilon_\rho$ and $\epsilon_z$, we have replaced $R_e$ by the mean radius $R$ because $R_e$ is not constant. The change thus introduced is of second-order with respect to the flattenings and would matter only if second-order terms were included in the equation of $\rho$.}. We note that $\rho$ is a known time function when the orbit of the companion around the primary is known. 

The proportionality constant $\gamma$ (a.k.a. relaxation factor) is a radial deformation rate gradient and has dimension $T^{-1}$. It is worth mentioning that Eq. (\ref{eq:creep}) is obtained from the integration of a spherical approximation of the Navier-Stokes equation of a radial flow across the surface of the ellipsoid in the case of a very-low-Reynolds-number flow (Stokes flow). In this approximation, the inertia tensor can be neglected and the stress due to the non-equilibrium is included in the pressure term. The solution of this approximation compared to Eq. (\ref{eq:creep}) shows that 
\beq\label{eq:gamma}
\gamma \simeq \frac{wR}{2\eta} \simeq \frac {3gm}{8\pi R^2 \eta},
\endeq
where $g$ is the gravity acceleration at the surface of the body, $R$ is the mean radius, $\eta$ the viscosity and $w$ the specific weight at the surface of the body (see Appendix A).

Often, the system formed by the primary and the companion is isolated and their relative motion is Keplerian. For short time intervals, the orbital evolution can be neglected and, at each instant, the relative position of the companion is fixed by 2 quantities: the radius vector $r(t)$ and the true longitude $\varphi(t)$, which are both known functions of the time. It is important to emphasize that the angles $\varphi$ and $\widehat\varphi$ may be referred to the same origin, and so the orbital longitude also includes the rotation angle $-\Omega t$. \footnote{The alternative choice of adding $\Omega t$ to the longitude  $\widehat\varphi$ is also possible (see \cite{Rh1}). }

The creep equation may be written as
\beq\label{eq:edo}
\dot\zeta + \gamma\zeta = \gamma\rho(t).
\endeq
We thus have a first-order linear differential equation to be solved. The general solution is 
\beq\label{eq:zetaint}
\zeta=e^{-\gamma t} \int_t \gamma \rho(t) e^{\gamma t} dt.
\endeq
One difficulty with this approach is that we have implicitly assumed that $\Omega$ is a constant. This is approximately true when the body is in free rotation, but when  
the primary is trapped into a synchronous motion, the frequency $\Omega-n$ is close to zero and the forced oscillation of $\Omega$ can no longer be neglected. In such case Eq. (\ref {eq:zetaint}) cannot be used.
  
\section{The equations of Folonier et al. \cite{Rh3}}

The lower-order approximation of a smooth function over a sphere is the triaxial ellipsoid. If we impose that the center of the ellipsoid coincides with the center of gravity of the body, we may approximate the solution of Eq. (\ref{eq:edo}) as
\begin{eqnarray} 
\zeta(\widehat{\theta},\widehat{\varphi},t) &=& R_e\left(1+\frac{1}{2}\mathcal{E}_\rho \sin^2{\widehat{\theta}}\cos{(2\widehat{\varphi}-2\varphi_\mathcal{B})}-\mathcal{E}_z \cos^2{\widehat{\theta}}\right),
\label{eq:zeta}
\end{eqnarray}
where the instantaneous flattenings $\mathcal{E}_\rho$, $\mathcal{E}_z$ and the lag  of the bulge $\varphi_\mathcal{B}=\varphi+\delta$ are 
unknown functions of the time. Here, $\delta$ is the lag of the bulge vertex with respect to \tens{M} (Fig. \ref{fig:delay}).

Before proceeding, we may remind that the solution of Eq. (\ref{eq:edo}) in \cite{Rh1, Rh2} was given by a sum of ellipsoidal bulges over one sphere of radius $R$, each of them with its own flattenings and lag. However, to the first order of approximation, the sum of two or more ellipsoidal bulges is one ellipsoidal bulge with its own flattenings and lag. 

If this approximation for $\zeta$ is substituted into Eq. (\ref{eq:edo})\footnote{We use the approximation $R_e \simeq R(1+\frac{1}{3}\epsilon_z)$ to introduce the constant mean radius $R$ in the equations}, we obtain
 
\begin{eqnarray}
&&\left(\Big(\dot{\mathcal{E}}_\rho + \gamma\mathcal{E}_\rho\Big)\cos{2\delta}+\mathcal{E}_\rho(2\Omega-2\dot\varphi-2\dot{\delta})\sin{2\delta}\right)\frac{1}{2}  \sin^2{\widehat{\theta}}\cos{(2\widehat{\varphi}-2\varphi)}\\
&&+\left(-\mathcal{E}_\rho(2\Omega-2\dot\varphi-2\dot{\delta})\cos{2\delta}+\Big(\dot{\mathcal{E}}_\rho + \gamma\mathcal{E}_\rho\Big)\sin{2\delta}\right)\frac{1}{2}  \sin^2{\widehat{\theta}}\sin{(2\widehat{\varphi}-2\varphi)}\nonumber\\
&&+\left(\dot{\mathcal{E}}_z+\gamma\mathcal{E}_z\right)\left(\frac{1}{3}-\cos^2{\widehat{\theta}}\right)= \frac{1}{2}  \sin^2{\widehat{\theta}}\gamma\epsilon_\rho\cos{(2\widehat{\varphi}-2\varphi)}+\gamma\epsilon_z\left(\frac{1}{3}-\cos^2{\widehat{\theta}}\right).\nonumber
\end{eqnarray}

Since the flattenings $\mathcal{E}_\rho,\mathcal{E}_z$ and the lag angle $\delta$ cannot depend on the coordinates $\widehat{\theta},\widehat{\varphi}$, the coefficients of independent trigonometric functions of the coordinates in the above equation may be satisfied separately and the equation may be split into three equations which must be satisfied separately. They are
\begin{eqnarray}
\dot{\delta}          &=& \Omega-\dot{\varphi}-\frac{\gamma\epsilon_\rho}{2\mathcal{E}_\rho} \sin{2\delta} \nonumber\\
\dot{\mathcal{E}}_\rho &=& \gamma\Big(\epsilon_\rho \cos{2\delta}-\mathcal{E}_\rho\Big) \nonumber\\
\dot{\mathcal{E}}_z   &=& \gamma\Big(\epsilon_z-\mathcal{E}_z\Big).
\label{eq:Folo3}
\end{eqnarray}
This system of differential equations of first order allows us to calculate the time evolution of the instantaneous flattenings and the lag angle, when the orbital motion of the companion (that is, $r(t)$ and $\dot{\varphi}(t)$) and the spin rate $\Omega(t)$ are known. 

For analytical studies, it is convenient to transform the equatorial prolateness ${\mathcal{E}}_\rho$ and the lag angle $\delta$ in their Cartesian counterparts
\begin{eqnarray}
x & = & \frac{{\mathcal{E}}_\rho}{\overline\epsilon_\rho} \cos 2\delta \\ \nonumber
y & = & \frac{{\mathcal{E}}_\rho}{\overline\epsilon_\rho} \sin 2\delta \\ 
\end{eqnarray}
where
\beq
{\overline\epsilon_\rho} = {\epsilon_\rho}\left( \frac{r}{a}\right)^3=
\frac{15}{4}\left(\frac{M}{m}\right)\left(\frac{R}{a}\right)^3.
\endeq
The two first equations then become 
\begin{eqnarray}
\dot{x} & = & -\gamma x - \nu(t) y + \gamma \left( \frac{a}{r}\right)^3 \\ \nonumber
\dot{y} & = &  \nu(t) x - \gamma y 
\end{eqnarray}
where
\beq
\nu(t)=2\Omega - 2\dot\varphi,
\endeq
or, using the complex variable $Z=x+iy$,
\beq
\dot{Z} + \big( \gamma - i \nu(t) \big) Z = 
\gamma \left( \frac{a}{r}\right)^3 .
\endeq 
The third equation, for ${\mathcal{E}}_z$, can be considered separately.

In the approximated case in which $\Omega$ is assumed to be constant, this system is a linear differential equation whose solution is
\beq
Z=Ke^{-\int{f(t)dt}} + \gamma e^{-\int{f(t)dt}} \int e^{\int{f(t)dt}} \left(\frac{a}{r}\right)^3 dt 
\endeq
where
\beq
f(t) = \gamma - i \nu(t) 
\endeq
and $K$ is a complex integration constant (see\cite{Arf} Sec 9.2). In this approximation, this solution is the same obtained for eqn. (\ref{eq:edo}) in \cite{Rh2}. 

In the general case, when the variation of $\Omega$ must be taken into account, we need one more equation for $\dot{\Omega}$. In order to obtain this additional equation, we need to know the torque acting on the primary. It is
\beq
\dot\Omega=-\frac{M_z}{C}
\endeq
where $C$ is the polar moment of inertia of the primary and $-M_z$ is the reaction to the torque of the disturbing force acting on the companion (see Appendix B). Hence
\beq\label{eq:Folo4}
\dot\Omega = -\frac{3GM{\mathcal{E}}_\rho}{2r^3}\sin 2\delta  
= -\frac{3GM{\overline{\epsilon}_\rho}}{2r^3}y 
\endeq
where we discarded higher-order terms and the term $\dot{C}\Omega/C$ (since $\dot{C}$ is proportional to ${\mathcal{E}}_z$).  

This new version of the creep equations is formally analogous to the equations obtained by Correia et al. \cite{CBLR}  in the tidal theory based on the Maxwell viscoelastic model. In fact, that theory and the creep tide theory become equal when the elastic terms are neglected \cite{SFM-AA}. 

\section{Rotation and Dynamical Equilibrium Figure}\label{sec:Rot}

The integration of the system formed by Eqs. (\ref{eq:Folo3}) and (\ref{eq:Folo4}) allows us to know the evolution of the equilibrium figure and rotation of the primary body. We may start with the simple case of circular orbits in the constant rotation approximation. In this case the equation for $Z$ becomes
\beq
\dot{Z} + (\gamma - i \nu) Z =\gamma
\endeq
where the semi-diurnal frequency $\nu$ is assumed as constant. 
The solution of this equation is trivial:
\beq
Z = Ce^{-\gamma t}e^{i\nu t} + \frac{\gamma}{\gamma-i\nu}
\endeq
where $C$ is an integration constant.
If we discard the free component because it is transient and tends exponentially to zero, the solution is reduced to the complex constant $\displaystyle\frac{\gamma}{\gamma-i\nu}$, that is to
\begin{eqnarray}
\delta &=&\half \arctan \left(\frac{\nu}{\gamma}\right) \\
E_\rho \defeq \frac{{\mathcal{E}}_\rho} {\overline{\epsilon}_\rho} &=& \frac{\gamma} {\sqrt{\gamma^2+\nu^2}} = 
\cos 2\delta.
\end{eqnarray}
The extreme cases $\gamma\gg \nu$ and $\gamma\ll \nu$ are worth being discussed separately as they correspond to the great majority of the known systems. They characterize two types of bodies that respond to tidal forces in very different ways: gaseous bodies and stiff bodies. 

{Gaseous bodies} have relaxation factors generally of order $1-100 {\rm\ s}^{-1}$ (\textit{cf}. \cite{Rh1}, table 1). Thus, normally, $\gamma \gg \nu$ and $\delta$ is very small.  In this case, the solutions are the same as in Darwin theory, i.e. $\delta$ is a very small quantity proportional to the semi-diurnal frequency $\nu$ (the so-called CTL theories) and the equatorial prolateness is close to the hydrostatic value $\overline{\epsilon}_\rho$.

{Stiff bodies} have very small relaxation factors, less than $10^{-6} {\rm\ s}^{-1}$ (\textit{cf}. \cite{Rh1}, table 1),  and $\gamma \ll \nu$ except in the close neighborhood of the synchronous motions (in which case, $\nu\sim 0$).  
In this case $\delta \sim 45^\circ$ and $E_\rho$ tends to zero. The lag close to $45^\circ$ is striking because all classical theories assume that $\delta$ is always small. However, the most significant result is that $E_\rho=\gamma/\nu \sim 0$.This means that the tidal forces in this case are not enough strong to create a significant deformation in the body. One typical example is Mercury, in which case $\gamma$ is so small that the tidal forces are currently not able to affect the shape of the body and the observed equatorial prolateness is several orders of magnitude larger than the prolateness that could be expected as due to the tidal forces on the planet (for a discussion, see \cite{Gom}). 

The solution of the third of Eqns. (\ref{eq:Folo3}), in the circular approximation is also trivial:
\beq
\mathcal{E}_z= C'e^{-\gamma t} + \overline{\epsilon}_z, 
\endeq 
where $C'$ is an integration constant and  
\beq
\overline\epsilon_z=\half\ \overline\epsilon_\rho + \frac{5R^3\Omega^2}{4mG}.
\endeq

\subsection{General examples}

In the case of eccentric orbits, the analytical integration becomes cumbersome as the Keplerian variables are given by power series, but it is still feasible when the variations in the rotation speed may be discarded. In the full case, with variable $\Omega$, the solutions can only be obtained via numerical simulations. The results depend on the nature of the body (stiff or gaseous) and on the type of motion (free rotation or resonant). Some typical examples with $e=0.2$ are given below.

\begin{figure}[t!]
	\begin{center}
\resizebox{0.890\columnwidth}{!}{%
  \includegraphics{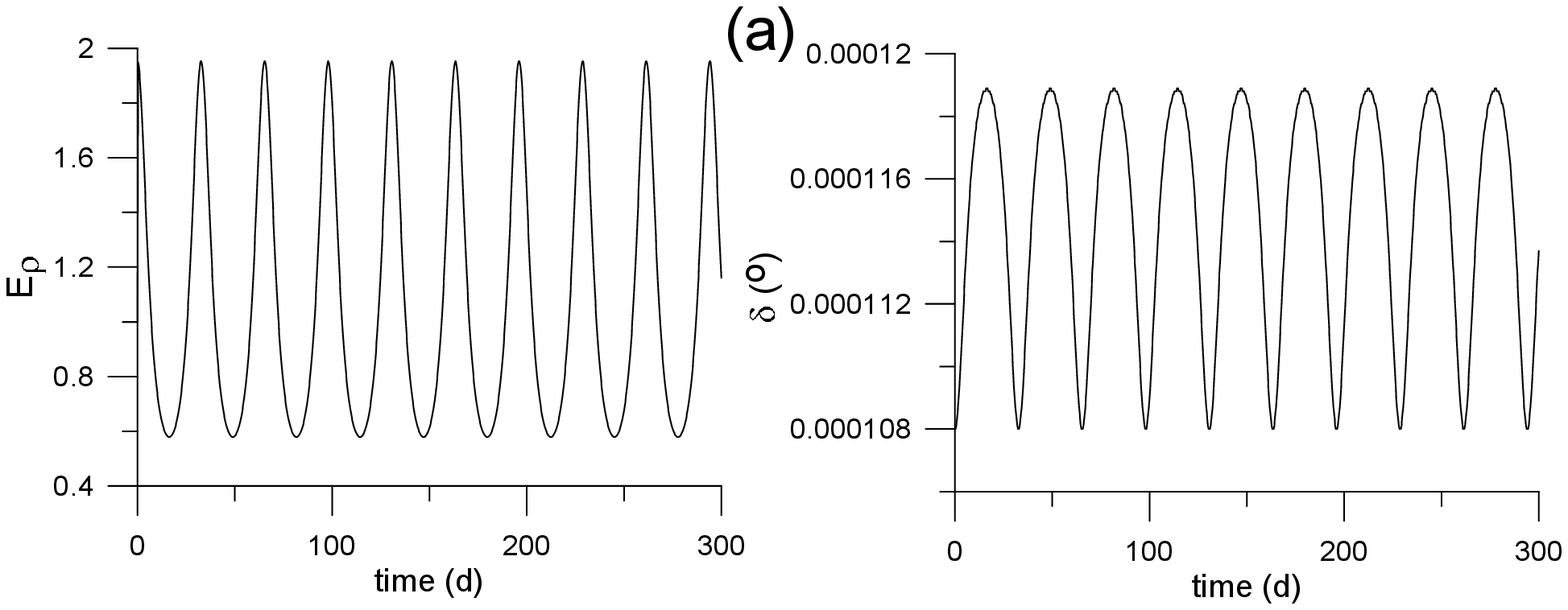} }
\resizebox{0.890\columnwidth}{!}{%
	\includegraphics{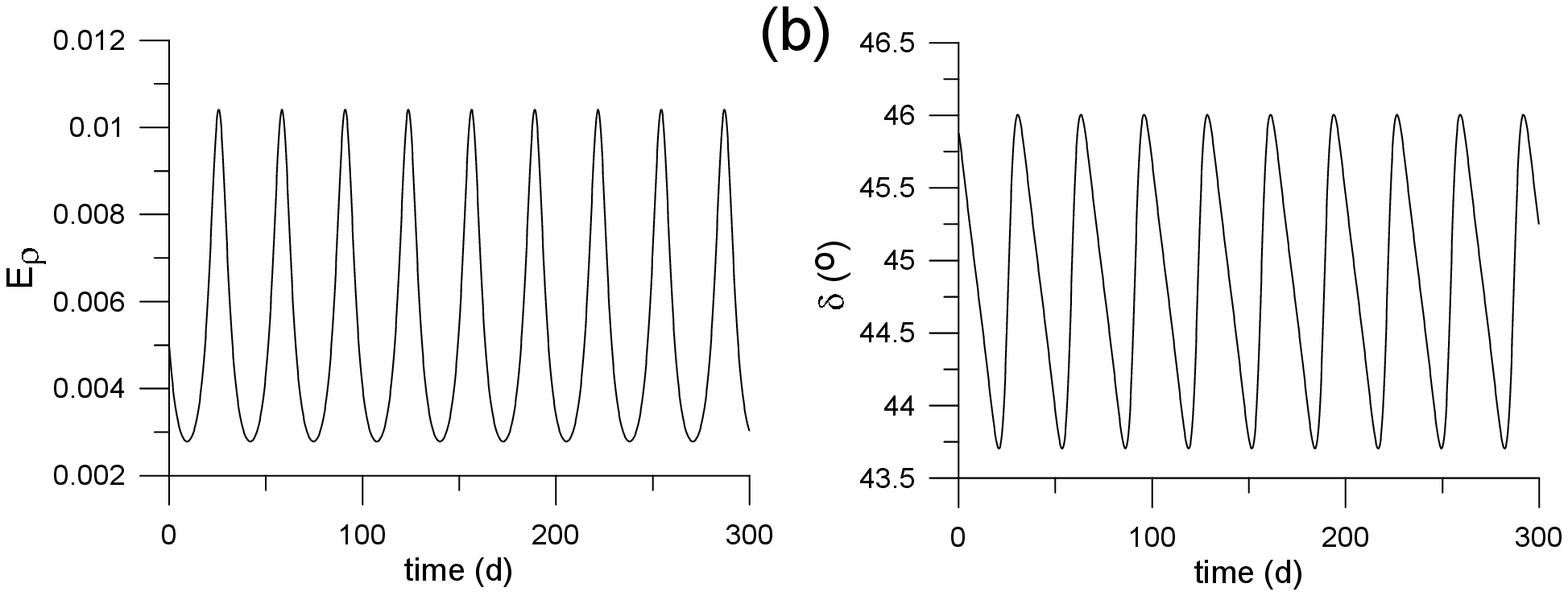} }
\resizebox{0.890\columnwidth}{!}{%
	\includegraphics{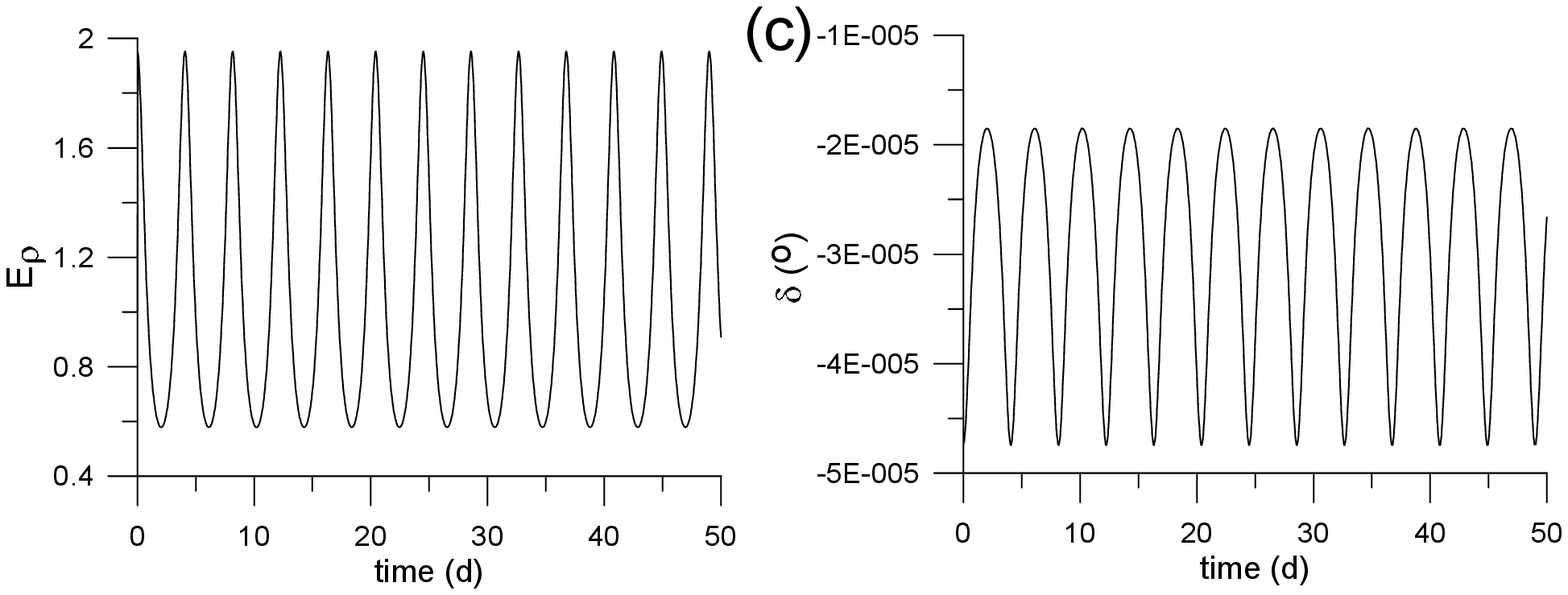} }
\resizebox{0.890\columnwidth}{!}{%
	\includegraphics{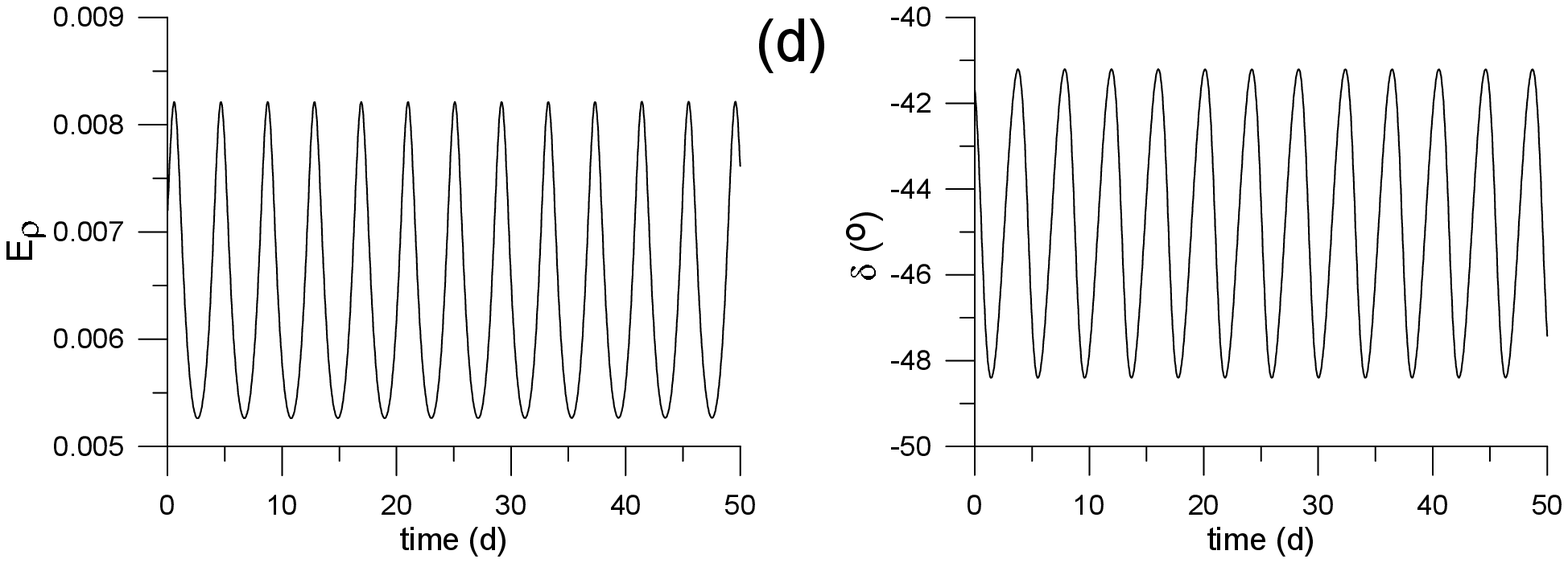} }
\caption{Variations of $E_\rho$ and $\delta$ in the case of rotating bodies. From top to bottom: (a.) Neptune in a 33-day orbit. (b.) Earth in a 33-day orbit. (c.) Slow rotating star hosting a Jupiter in a 4-day orbit. (d.) Slow rotating Earth in a 4-day orbit. In all cases $e=0.2$.}
		\label{fig:rot}
	\end{center}
\end{figure}  

\subsubsection{Fast rotating Neptune in a 33-day orbit}\label{sec:NR}

In this example, we consider one exoplanet with the same physical characteristics as Neptune, in a 33-day orbit (i.e. $a=0.2$ AU). The considered relaxation factor is $\gamma = 10\ {\rm s}^{-1}$ and the planet is assumed to have a fast rotation, $P=3.27$ d. The results in fig. \ref{fig:rot} (a) show that the lag $\delta$ is very close to 0 (as in Darwin's theory), and the equatorial prolateness is of the order of the prolateness of the Jeans ellipsoid, that is, the hydrostatic equilibrium, expected in a circular motion. The minimum $\delta$ and the maximum prolateness are reached when the planet is at the pericenter of its orbit. Because of the large eccentricity adopted, the height of the tidal bulge may become up to 2 times larger than the tidal bulge of the circular case. The rotation velocity remains constant during the whole simulation (relative variation of less than $10^{-8}$ in one year).   
   
\subsubsection{Fast rotating Earth in a 33-day orbit}\label{sec:TR}

This example is very similar to the previous one, with the same orbital and rotational parameters. However, the physical characteristics of the planet are similar to the Earth's ones. The considered relaxation factor is $\gamma = 2 \times 10^{-7}\ {\rm s}^{-1}$. The results in fig. \ref{fig:rot} (b) show that the lag $\delta$ is very close to 45 degrees (at variance with the small lag assumed in Darwin's theory), and the equatorial prolateness due to the tide is very small. Even in the more favorable condition, near the pericenter, it remains of the order of one hundredth of the prolateness of the corresponding Jeans ellipsoid. The maximums and minimums of $\delta$ occur when the planet is at $\sim 74$ degrees away from the pericenter, resp., approaching or going away from it. The rotation velocity remains constant during the whole simulation (relative variation of about $10^{-6}$ in one year).   
    
\subsubsection{Slow rotating central star}\label{sec:SR}

In this example, we consider one Sun-like star hosting a hot-Jupiter in a 4-day orbit (i.e. $a=0.05$ AU). The considered relaxation factor is $\gamma = 50\ {\rm s}^{-1}$ and it is assumed to have a slow rotation, $P=30$ d. The results in fig. \ref{fig:rot} (c) show that the equatorial prolateness due to the tide is very similar to that of a fast rotating gaseous planet as the one discussed in section \ref{sec:NR}. The lag $\delta$ is very close to 0 (as in Darwin's theory) but, at variance with the previous cases, it is negative. This point shall be stressed to avoid hasty applications of the tidal formulas used to study planetary satellites, in the study of exoplanets. If the central body (star) rotates slowly, $\delta$ is negative and all tidal effects (including those on the orbit of the planet) are reversed. 
The minimum $\delta$ (maximum in absolute value) and the maximum prolateness are reached when the planet is at the pericenter of its orbit. As in the studied case of a fast Neptune, the height of the tidal bulge may become up to 2 times larger than the tidal bulge expected in the circular case. The rotation velocity is increasing, but the interval considered is too short to make visible any variation. It is important to remind that the rotation of solar-type stars is not ruled only by tidal effects; the stellar braking of the star due to the loss of mass and angular momentum  via stellar winds is important and in the early life of the star may completely dominate the rotational evolution of the star (see \cite{host}). 

\subsubsection{Slow rotating Earth in a 4-day orbit}\label{sec:TL}

An example like the previous one with a stiff primary does not exist. In order to know the dynamical figure of a slow rotating stiff body, we consider one slow rotating Earth in a 4-day orbit around a Sun-like star (i.e. $a=0.05$ AU). The considered relaxation factor is $\gamma = 2 \times 10^{-7}\ {\rm s}^{-1}$ and the planet is assumed to have a slow rotation, $P \sim 35$ d. The results in fig. \ref{fig:rot} (d) are similar to those of the fast rotating Earth with the difference that, now, the lag $\delta$ is negative (close to $-45^\circ$ ). 

\begin{figure}[t]
	\begin{center}
		\resizebox{0.89\columnwidth}{!}{%
			\includegraphics{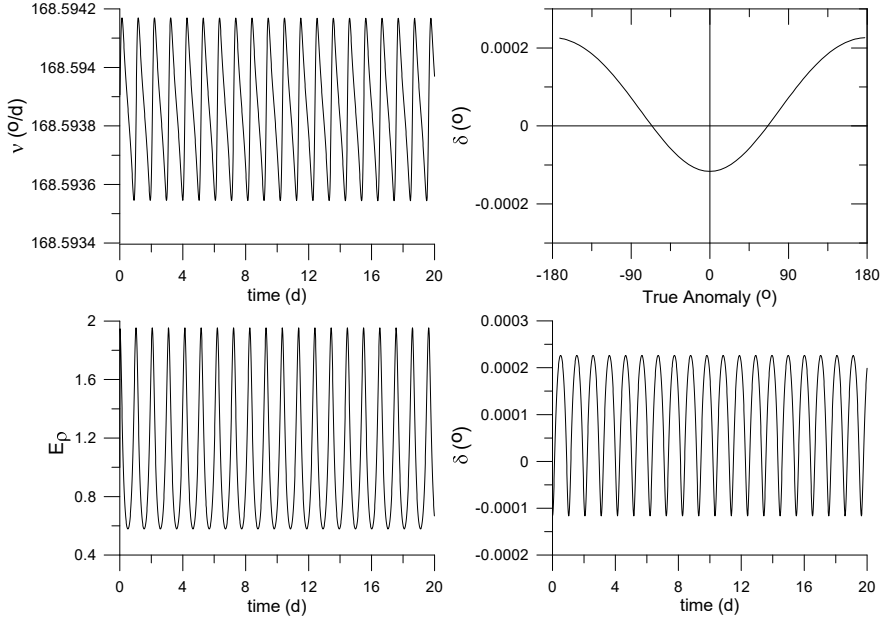} }
		\caption{Variations of $E_\rho$, $\delta$ and the semi-diurnal frequency $\nu$ in the case of one Neptune in stationary rotation. $a=0.02$ AU, $e=0.2$. (N.B. The rotation is faster than the synchronous one.) }
		\label{fig:NS}
	\end{center}
\end{figure}  
\begin{figure}[t]
	\begin{center}
		\resizebox{0.89\columnwidth}{!}{%
			\includegraphics{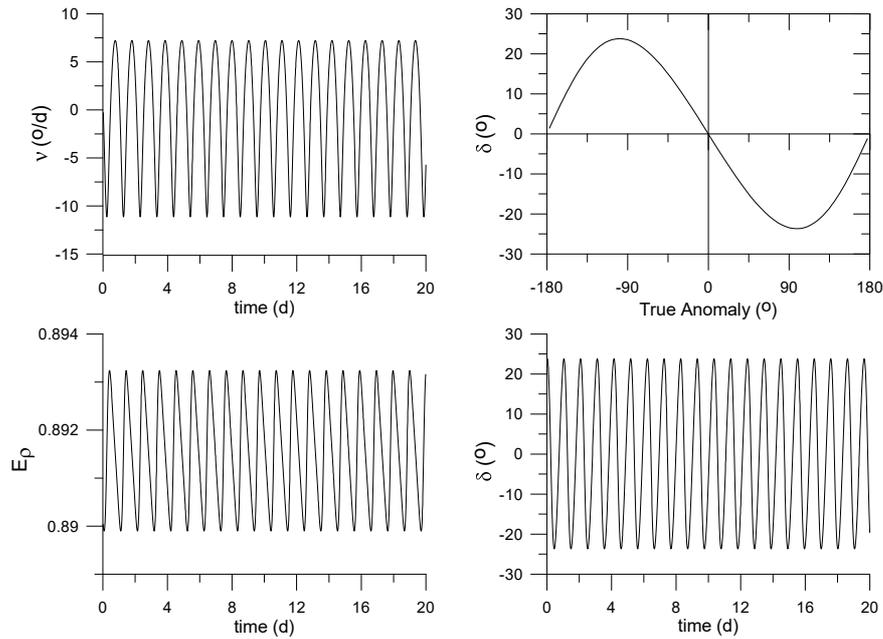} }
		\caption{Variations of $E_\rho$, $\delta$ and the semi-diurnal frequency $\nu$ in the case of an Earth in stationary rotation (with a forced oscillation about synchronism). $a=0.02$ AU, $e=0.2$.}
		\label{fig:TS}
	\end{center}
\end{figure}  

\subsubsection{Synchronous rotating Neptune in a 1-day orbit}

In this example, we consider one exoplanet with the same physical characteristics as Neptune, in a 1-day orbit ($a=0.02$ AU). The adopted relaxation factor is $\gamma = 10\ {\rm s}^{-1}$ and the planet's rotation is trapped in a stationary state $(\langle\dot{\Omega}\rangle = 0)$. The results in fig. \ref{fig:NS} show that the semi-diurnal frequency oscillates about a positive value. A simple calculation shows that this value is $ \nu \sim 12ne^2$ as predicted by the creep tide theory  for gaseous bodies \cite{Rh2} and by Darwin's CTL theory \cite{FRH}. The lag $\delta$ is very close to 0 and the equatorial prolateness is of the order of the prolateness of the Jeans ellipsoid, that is, the hydrostatic equilibrium. 
The maximum elongations of $|\delta|$ are reached at the pericenter and apocenter.
Because of the large eccentricity adopted, the height of the tidal bulge may become up to 2 times larger than that of the tidal bulge of the circular case. The rotational velocity is almost constant with just an oscillation of the order $10^{-6}$ of its absolute value.

\subsubsection{Synchronous rotating Earth in a 1-day orbit}

This example is very similar to the previous one, but the physical characteristics of the planet are similar to the Earth's ones. The considered relaxation factor is $\gamma = 2 \times 10^{-7}\ {\rm s}^{-1}$ and the planet's rotation is trapped in a stationary state $(\langle\dot{\Omega}\rangle = 0)$. The results in fig. \ref{fig:TS} show that the semi-diurnal frequency oscillates about a negative value close to zero, that is the stationary rotation of stiff bodies is almost synchronous and is not faster than the synchronous rotation as indicated  by Darwinian CTL theories \cite{FRH}. The lag $\delta$ has large elongations ($\sim 24^\circ$) around 0 and vanishes near the pericenter and apocenter. The height of the tidal bulge is smaller than that of the circular case. 

The rotation velocity is almost synchronous (i.e. $\Omega \sim n$), but is not constant. It shows a well-defined oscillation. This oscillation of the semi-diurnal frequency $\nu$ has been observed in several planetary satellites. One example is Mimas for which Cassini's images allowed the detection of a significant forced libration amplitude: $ 0.838^\circ \pm 0.002 ^\circ $ \cite{Taj}.  

\begin{figure}[t]
	\begin{center}
		\resizebox{0.9\columnwidth}{!}{%
			\includegraphics{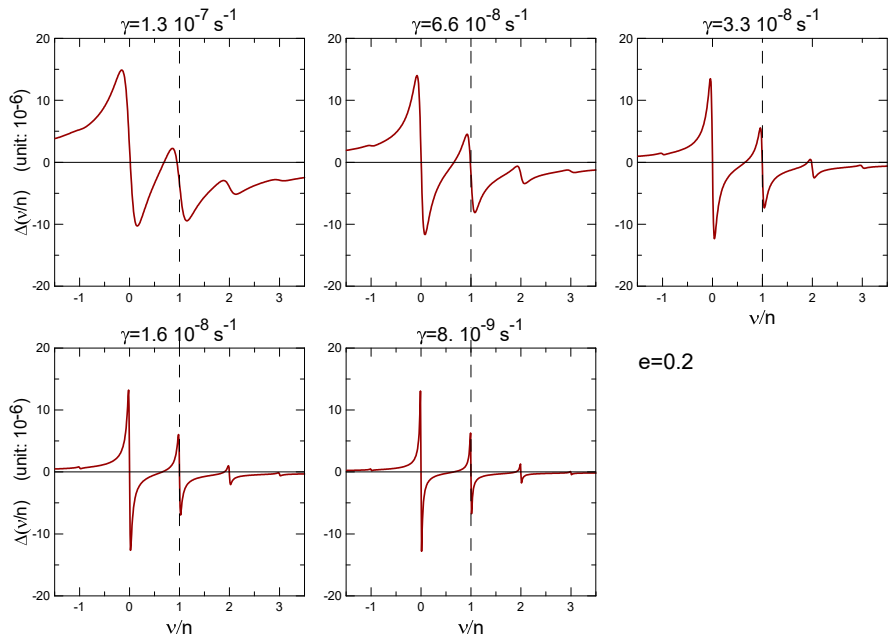} }
		\caption{Maps showing the variation of $\nu/n$ per period for several values of $\gamma$ in the case of one non-rigid Mercury with $e=0.2$ and $ n\simeq 8.27 \times 10^{-7}\ {\rm s}^{-1}$. Adapted from \cite{Rh2}.}
		\label{fig:Plot10}
	\end{center}
\end{figure}  

\subsubsection{Planets in spin-orbit resonance}

The evolution of the rotation of one body submitted to tidal torques due to an orbiting companion is ruled by Eqn. (\ref{eq:Folo4}). 
Thus, $\dot\Omega > 0$ (the rotation is accelerated) when $\sin 2\delta < 0$ and $\dot\Omega < 0$ (the rotation is slowed) when $\sin 2\delta > 0$. In general,  this means that the rotation evolves towards the synchronization of both motions, but the stationary motions reached are often not a true synchronization. For instance, in the case of gaseous bodies, the stationary solution is synchronous only when $e=0$. If $e \ne 0$, the stationary solution is such that the rotation of the body is faster than the orbital motion of the companion (i.e, it is supersynchronous). The condition $\langle \dot\Omega \rangle = 0$, is reached when $\Omega \simeq n(1+6e^2)$ \cite{Hut, FRH}. In the case of stiff bodies, however, the possibilities are more complex. The function  $\langle \dot{\Omega}\rangle$ has a stable zero very close to $\Omega=n$ (synchronization), but it may also have, for $e \ne 0$, several other stable solutions near some spin-orbit resonances as shown in fig. \ref{fig:Plot10}. The main one is seen at $\nu/n = 1$ (i.e $\Omega/n= \frac{3}{2})$ at the point where a descending branch of the function intersects the horizontal axis. It appears in all panels of fig.  \ref{fig:Plot10}, but may disappear either if we adopt larger values of $\gamma$ (namely, if $\gamma > 1.38 \times 10^{-7}\ {\rm s}^{-1}$ in the case of the figure shown) or smaller values of $e$. In these cases, the top of the kink of the curves at  $\nu/n = 1$ remains below 0. The rotation of the body, in this case, will decrease continuously without the possibility of trapping of the rotation at that spin-orbit resonance. The same can be seen at  $\nu/n = 2$ (i.e. $\Omega=2n$). The kink remains below 0 in the two first panels, but crosses the horizontal axis in the other panels creating a stable stationary solution. Looking closely, we may see other small kinks at $\nu/n=-1$ and $\nu/n=3$ (i.e. $\Omega/n= \frac{1}{2}$ and $\Omega/n= \frac{5}{2}$). If the eccentricity is larger, they will give rise to other spin-orbit resonances for low values of $\gamma$. It is noteworthy that the kinks appear at frequencies in the sequence  $\{\frac{k}{2} \}, (k=-1,0,1,2,\cdots)$ \cite{MakE, CBLR, Rh2}. Spin-orbit resonances of tidal origin at other frequencies do not exist.
\begin{figure}[t]
	\begin{center}
		\resizebox{0.9\columnwidth}{!}{%
			\includegraphics{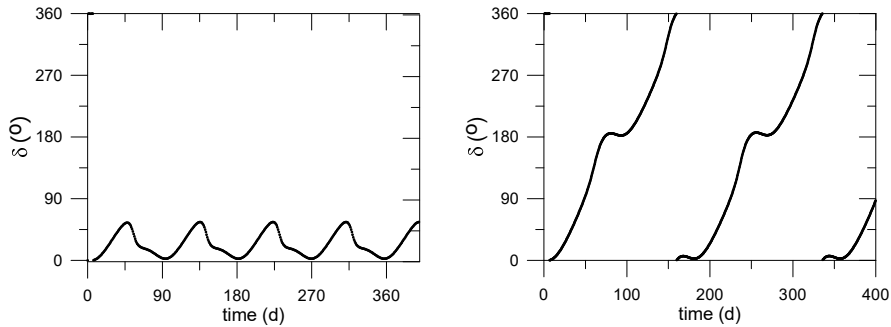} }
		\caption{Variation of $\delta$ in the case of a non-rigid Mercury crossing the 3/2 spin-orbit resonance (Period $\sim$ 58 days) for two different values of $\gamma$. Left: $\gamma= 6\times 10^{-7}\ {\rm s}^{-1}$, Right: $\gamma= 3\times 10^{-7}\ {\rm s}^{-1}$. Eccentricity: $e=0.2$.}
		\label{fig:M12}
	\end{center}
\end{figure}  

Fig. \ref{fig:M12} shows the evolution of the tidal bulge of one non-rigid Mercury-like planet whose rotation is in the 3/2 spin-orbit resonance, in two cases where the system is crossing the resonance, but in which $\gamma$ is not small enough to allow the system to be trapped. In the first case (left panel), the bulge continuously follows the Sun with a variable lag of some tens of degrees. In the second case (right panel), the body is stiffer and the bulge does not follow the Sun closely. The lag increases with time and the bulge is driven by the body rotation returning to the initial position in alignment with the companion only after 2 complete periods of the planet around it \footnote {To get the complete picture one has to remember that there are two opposite bulges}. When the planet is trapped in the 3/2 spin-orbit resonance, the behavior is like the one shown in this second case. In the case of a 2/1 spin-orbit resonance, one more possibility appears in which the bulge is also driven by the rotation of the body, but returns to the initial position after 1 complete period of the planet around the star \cite{Gom}.

\section{Energy dissipation}

In all papers on the creep tide theory, the bulk dissipation was calculated from the estimation of the total mechanical energy lost by the system. This approach is physically very simple. If the companion body is considered as a mass point (and this can be done when we are only considering the tides raised on the primary body), the energy tidally dissipated in the primary may only be originated in its rotation and on the companion's orbit. No other non-primeval source of energy exists able to continuously supply the tidal energy dissipated by the system. In this sense, the secular variation of the primary's rotation and of the semi-major axis of the relative orbit of the two bodies are the two gauges allowing us to evaluate the mechanical energy lost by the primary.

Since Darwin \cite{Darwin}, the problem of the tidal interaction of the two bodies is split into two parts. We consider first the deformation of the primary body due to an external body, the companion; then, we consider the perturbations of the motion of the companion, due to the disturbing potential due to this deformation. In all theories, the two steps are considered separately. 

In the first step, equilibrium equations, static or dynamic, are used to determine the deformation of the primary. This deformation changes the internal energy of the primary. The energy of the ellipsoid is \cite{Ess}
\beq
E_{\rm int} = -\frac{3Gm^2}{5R}\left(1-\frac{1}{15} \mathcal{E}_\rho^2-\frac{4}{45} \mathcal{E}_z^2 \right)
\endeq 
and its time variation is
\beq
\dot{E}_{\rm int} = 
\frac{2Gm^2}{25R}\left( \mathcal{E}_\rho \dot{\mathcal{E}}_\rho+\frac{4}{3} \mathcal{E}_z \dot{\mathcal{E}}_z \right).
\endeq 

In the second step, the forces acting on the companion due to the deformation of the primary are considered. This is a classical problem of dynamics as these forces are derivable from a time-dependent potential and its study can be done using classical Newtonian equations or the tools of Lagrangian and Hamiltonian dynamics. The variation in the orbital energy of the system is given by the time-derivative of the potential energy \cite{Rh3} 
\beq
\dot{E}_{\rm orb} = M\frac{\partial U}{\partial t} = M\sum_i \frac{\partial U} {\partial \lambda_i} \dot{\lambda}_i  
\endeq
where $\lambda_i$ are the three time-dependent parameters defining the deformed ellipsoid: the flattenings $\mathcal{E}_\rho, \mathcal{E}_z$ and the longitude of the equatorial bulge $\varphi_B$. $U$ is the potential given by Eqn. (\ref{eq:Ubis}). Hence, 
\beq
\dot{E}_{\rm orb}=-\frac{3GCM}{4r^3}\left(
\dot{\mathcal{E}}_\rho \cos 2\delta - 2\mathcal{E_\rho} (\dot\delta + \dot\varphi) \sin 2\delta  + \frac{2}{3} \dot{\mathcal{E}}_z \right).
\endeq

The second step is completed by the reaction of the torque $\mathbf{M}$ on the rotation of the primary. The rotational energy is $E_{\rm rot}= \frac{1}{2} C\Omega^2$, whose derivative, taking into account the value of $\dot{\Omega}$ given by Eqn. (\ref{eq:Folo4}) is
	
\beq 
\dot{E}_{\rm rot}= -\frac{3GCM\Omega}{2r^3}
\mathcal{E}_\rho  \sin 2\delta
+ \frac{2}{3}C \dot{\mathcal{E}}_z \Omega^2
\endeq
where we have taken into account not only the variation of $\Omega$ but also the variation of the polar moment of inertia: 
	\beq
	\dot{C}=\frac{2}{3}C \dot{\mathcal{E}}_z.
	\endeq

If these contributions are collected, we obtain the total variation of the mechanical energy of the system:
\beq
\dot{E}=-\frac{3GCM\gamma\epsilon_\rho}{4r^3}\sin^2 2\delta 
- \frac{2Gm^2}{25\gamma R}\left( \dot{\mathcal{E}}_\rho^2 + \frac{4}{3} \dot{\mathcal{E}}_z^2 \right). \label{eq:Edot}
\endeq
The first part of the above equation is immediately obtained when $\dot\delta$, in the equation for the variation of the orbital energy, is substituted by the corresponding Folonier equation. The second part comes from a more deep manipulation of the other terms in which not only all Folonier equations, but also the definitions of the moment of inertia of the homogeneous primary $C$ and the static flattenings $\epsilon_\rho, \epsilon_z$ are used. 

It is important to note that $\dot{E} < 0$, always, as expected. The mechanical energy always decreases and the decrement must correspond to the amount of tidal energy released inside the primary body.      
In our earlier papers, we have considered the work done by the perturbing forces (instead of the variation of the orbital energy) and the variation of the rotational energy. That setting is incomplete and the results differ from the one given here by some large short-period terms. However, both results become equal when they are averaged over the orbital period.

In the circular approximation (see sec. \ref{sec:Rot}), the forced components of $\dot{\mathcal{E}}_\rho, \dot{\mathcal{E}}_z$ vanish. So, the contributions associated with these components decrease exponentially to zero. The dissipated power (or dissipation rate) is then given by the first part of eqn. (\ref{eq:Edot}). In the circular approximation it may be written as

\beq
\dot{E}_{\rm circ}=-\frac{3GCM\nu}{4a^3} \
\overline\epsilon_\rho \frac{\nu\gamma}{\gamma^2+\nu^2}.
 \label{eq:Diss1}
\endeq

\begin{figure}[t]
	\centerline{\hbox{
			\includegraphics[height=3.5cm,clip=]{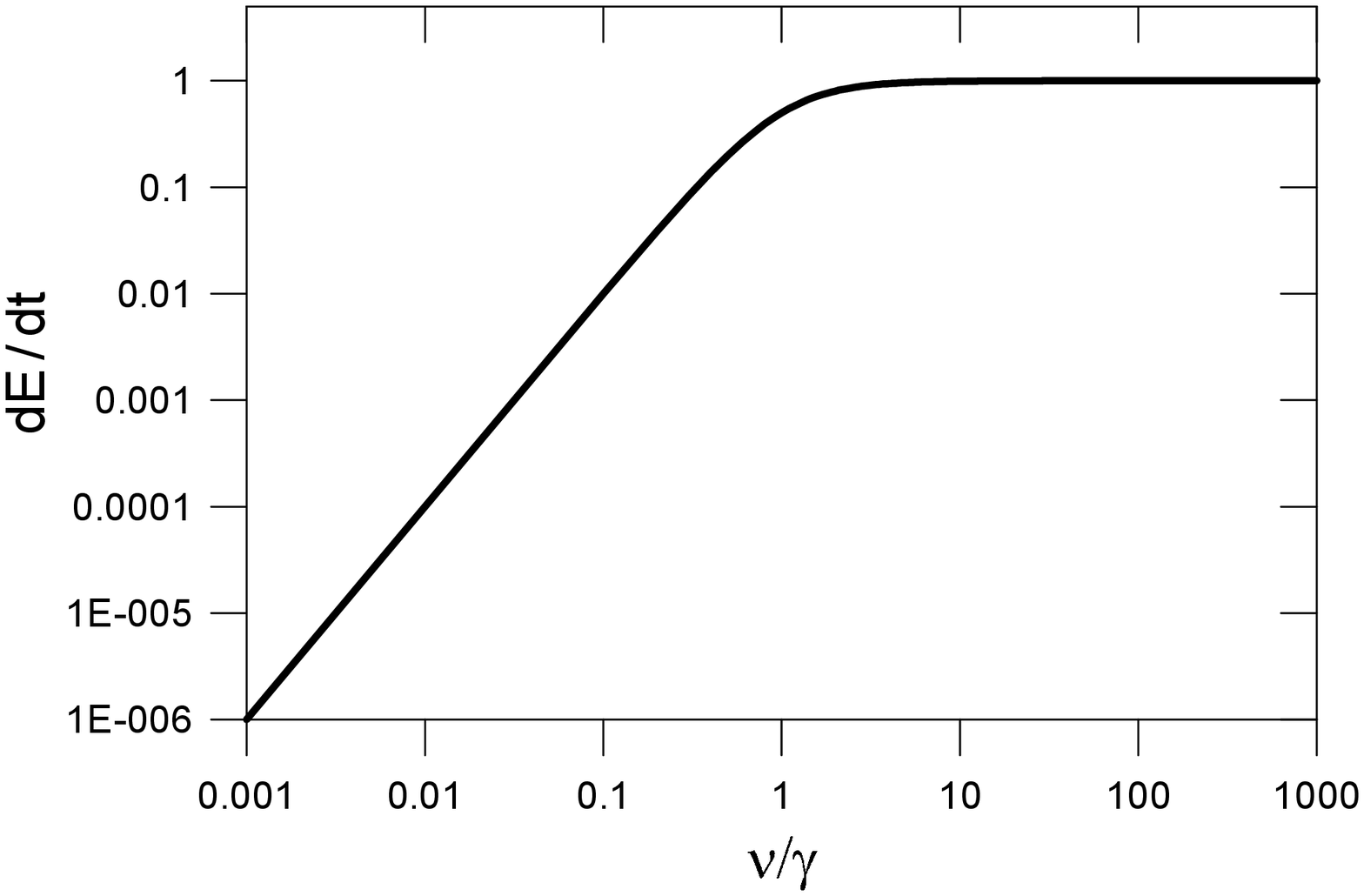}\hspace*{10mm}
			\includegraphics[height=3.5cm,clip=]{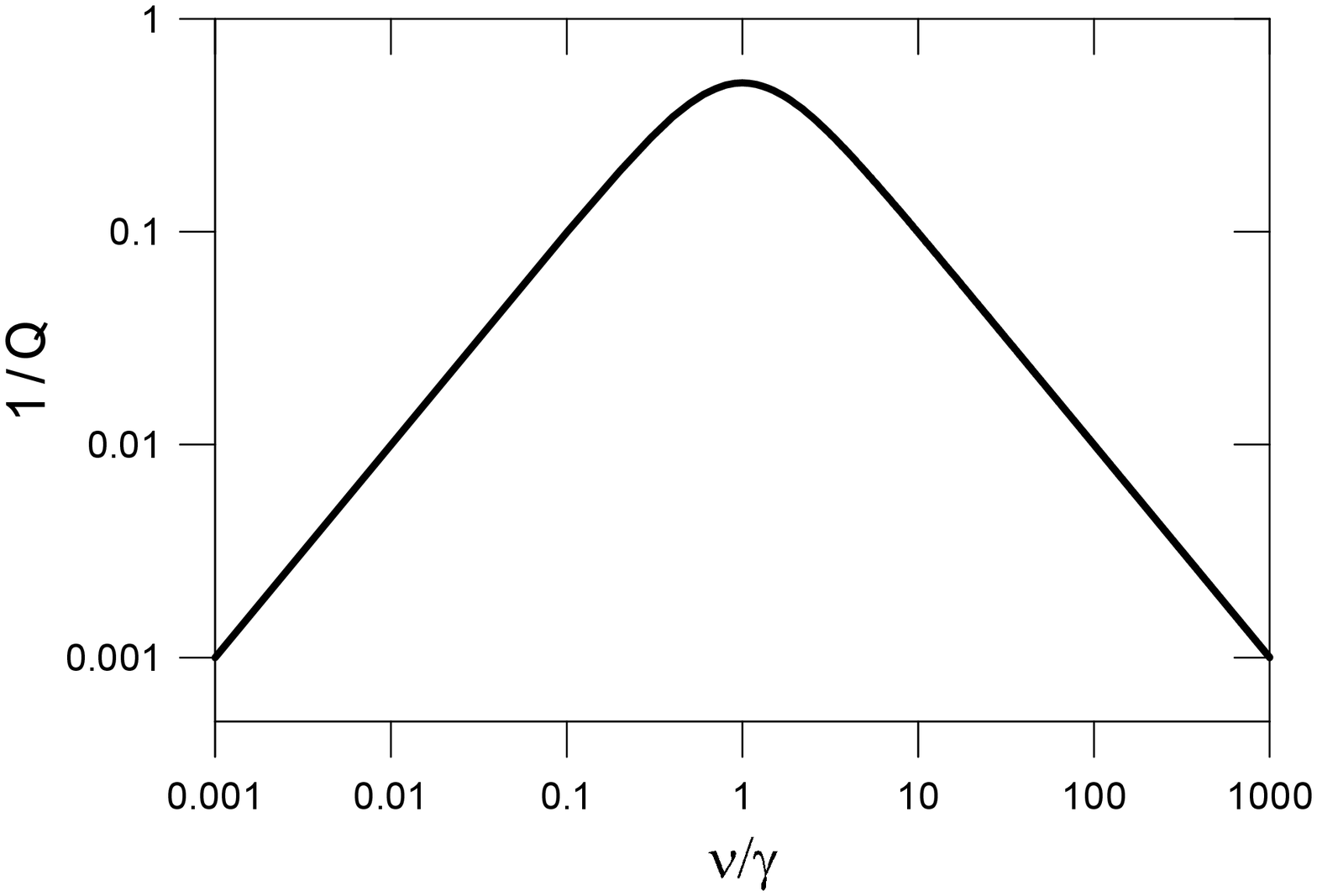}}}
	\caption{Dissipation in a free rotating body as a function of the frequency. Left: Dissipated power (in arbitrary units). Right: Equivalent $1/Q$.}
	\label{fig:Q}
\end{figure}

It is worth stressing the limit values obtained for the dissipated power in the two extreme cases: $\gamma \ll \nu$ (free rotating stiff bodies) and $\gamma \gg \nu$ (gaseous bodies).  
In the first case, we may neglect $\gamma$ in the denominator of the expression for $\dot{E}$ and the resulting expression for the dissipated power becomes independent of the semi-diurnal frequency. In the second case, we may neglect $\nu$ in that denominator and the resulting dissipated power becomes proportional to $\nu^2$ (see fig. \ref{fig:Q} left). This result has not been duly emphasized in previous papers. 
 
This quadratic dependence of the dissipated power with $\nu$ means that Eqn. (\ref{eq:Diss1}) is not sufficient to express the dissipated power in the neighborhood of the synchronization. In this case, a more elaborate calculation is necessary starting from the solutions of the equations for $\delta, \mathcal{E}_\rho, \mathcal{E}_z$ in Fourier series in the mean anomaly of the companion, and making, afterwards, an averaging over the short periods. We obtain, to the first order in the flattenings \cite{Rh3}:

\beq
\dot{E}_{\rm sync}=-\frac{21GCMne^2}{2a^3} \
\overline\epsilon_\rho \frac{n\gamma}{\gamma^2+n^2}.
\label{eq:Diss2}
\endeq

If we, purposely, forget Kepler's law and interpret $n$ and $a$ as independent variables, one carrying information on the frequency of the tide and the other on the distance of the primary to the companion, then, as before, the result becomes almost frequency independent when $\gamma \ll n$ and proportional to $n^2$ when $\gamma \gg n$.

If these results are applied to Enceladus, it is enough to assume that the satellite viscosity is consistent with the viscosity of melting ice ($10^{13} - 10^{14}$\ Pa s) to get for the tidal dissipation one value in the range of values determined from Cassini's observations (5-16 GW) \cite{Rh3}. The results for Mimas with a higher viscosity due to the frozen ice are also consistent with the negligible dissipation and the absence of observed tectonic activity.

\subsection {The quality factor} 

The quality factor $Q$ is a parameter that was first introduced in the study of the attenuation of free oscillations in electrical systems: the higher the $Q$, the better the system's ability to preserve its free oscillations undamped. It was later extended to quantify the loss of energy through forced oscillations of a non-linear system and introduced in tidal evolution theories \cite{Gol, Kaula, MacDo}. There, it was often defined as the inverse of the lag of the tide harmonic more influential in the dissipation (semi-diurnal or orbital), but a consensus on the exact law relating $Q$ and the lags does not exist \cite{Ef12b}.

The creep tide theory does not introduce ad hoc lags and, because of some vagueness in the definition of $Q$, there is no straightforward way of introducing it in this theory. It is however important to have conversion formulas allowing to relate $Q$ to the parameters used in creep tide models. 

The easiest way to do it is through the comparison of the expressions for the dissipated power obtained in both theories, when similar settings are adopted. The comparison of the equations  (\ref{eq:Diss1}), (\ref{eq:Diss2}) with the corresponding ones in Darwin's Theory (see \cite{FRH}, Eqns. 48 and 51) leads to the empirical relation
\beq
Q=\chi+\chi^{-1}
\label{eq:Q}
\endeq
where $\chi$ is the relaxation factor $\gamma$ in units of the frequency of the harmonic dominating the dissipation, that is $\chi=\gamma/\nu$ in the case of a free rotating body or $\chi=\gamma/n$ in the case of a body of synchronized rotation. Because of the symmetry of $\chi$ in Eq. (\ref{eq:Q}), the fraction defining it is some times inverted (e.g. in \cite{Rh1})

To complete the discussion started above, on the dissipation rate in the creep theory,  it is worth to mention that in classical Darwinian theories, the dissipation power is proportional to $\nu/Q$ (or $n/Q$). The quantity inversely proportional to $Q$ is not the dissipated power, but its integral over one complete cycle: $\Delta E=\displaystyle{\oint} \dot{E}dt$ (see \cite{Kaula} Eqns. 18-20). Fig. \ref{fig:Q} compares the dissipated power of a free rotating body in function of the frequency to the corresponding inverse of $Q$. When $\nu \gg \gamma$ the dissipated power no longer changes with the frequency. The decreasing branch of the curve $1/Q$, that corresponds to the behavior of a Maxwell viscoelastic body, does not mean that a change occurs in the body response when $\nu$ increases. $1/Q$ is proportional to the total energy dissipated in one cycle and is thus proportional to the decreasing period of the cycle.

\section{Orbital evolution}

The equations for the tidal evolution of the orbital elements due to the deformations of the primary are the same of the classical theories for the perturbations of the motion around one triaxial ellipsoid with given flattenings and orientation. However, the ellipsoid is not rigid and their parameters vary and may be calculated simultaneously with the integration of the given variational equations. This can be done by taking lags proportional to the frequency (as in Darwinian CTL theories), lags given by some inverse power law (as in the theories of Efroimsky and collaborators), by using the analytical solution of the creep equations \cite{Rh1, Rh2}, or by simultaneously integrating the Folonier et al. equations. 

We resume in this paper, for the sake of completeness and because of their usefulness, some results on the evolution of the orbital elements giving the size and the shape of the orbit: semi-major axis and eccentricity. They are given by
\beq
\dot{a}=\frac{2a^2}{GMm}\dot{W}
\label{eq:dota}
\endeq
\beq
\dot{e}=\frac{1-e^2}{e}\left(\frac{\dot{a}}{2a}-\frac{\dot{\mathcal L}}{\mathcal L} \right)
\label{eq:dote}
\endeq
where $\dot{W} = \mathbf{F}\cdot \mathbf{V}$ is the derivative of the work done by the disturbing force acting on the companion\footnote{Eqn. (\ref{eq:dota}) is obtained by derivation from the definition of the orbital energy: $E_{\rm orb} = -\frac{GMm}{2a} + M \delta U$ used instead of the classical expression of the energy of the Keplerian motion because of the explicit time dependence of $U$. In \cite{FRH, Rh1, Rh2}, this distinction was not done, but it is easy to check that the expression used was the same as here. } and $\mathcal L$ is the angular momentum. Eqns. (\ref{eq:dota}) and (\ref{eq:dote}) are the same equations that are obtained when the variational equations of Lagrange or Gauss are used (see \cite{Rh3}).

It is worth reminding that the equations given here consider only the deformations in one of the bodies (the primary). Often the tides in the two bodies contribute to the orbital evolution. The contributions due to the tides in the other body are given by the same equations as below, just inverting the roles played by the two bodies.   

\subsection{Free rotating bodies}
In this case, we generally use the approximation of the creep equations obtained neglecting the short period oscillations of $\Omega$. If the value of $\langle \dot{W} \rangle$ is taken from \cite{Rh2}, we obtain
\begdi
\langle \dot{a}\rangle = \frac{3nC \overline{\epsilon}_\rho}{ma}
\left((1-5e^2)\frac{\gamma\nu}{\gamma^2+\nu^2} 
-\frac{3e^2}{4}\frac{\gamma n}{\gamma^2+n^2} \hspace{3cm} \right.
\enddi
\beq \left. \hspace{2cm}
+\frac{e^2}{8}\frac{\gamma(\nu+n)}{\gamma^2+(\nu+n)^2}   
+\frac{147e^2}{8}\frac{\gamma(\nu-n)}{\gamma^2+(\nu-n)^2}
 \right) + \mathcal{O}(e^4). 
\endeq 
More compact equations may be obtained in particular cases where we may neglect $\gamma$ or $n$. 

For the eccentricity, we have \cite{Rh2} 
\begdi
\langle \dot{e}\rangle = -\frac{3nC \overline{\epsilon}_\rho e}{4ma^2}
\left( \frac{\gamma\nu}{\gamma^2+\nu^2} 
+ \frac{3}{2} \frac{\gamma n}{\gamma^2+n^2} \hspace{4cm} \right.
\enddi
\beq \left.  \hspace{2cm}
+\frac{1}{4}\frac{\gamma(\nu+n)}{\gamma^2+(\nu+n)^2}   
-\frac{49}{4}\frac{\gamma(\nu-n)}{\gamma^2+(\nu-n)^2} \right) 
+ \mathcal{O}(e^3). 
\endeq 
\noindent {\sl{Warning}}. 
Somewhere in \cite {Rh2} the Kepler law was used in the form $GM=n^2a^3$. That approximation is only valid when $m\ll M$. The equations given above were modified to be valid also in the general case.
   
\subsection{Bodies in stationary rotation}

In the case of synchronous or stationary supersynchronous rotation, it is necessaary to use solutions of the creep equations including the short period oscillations. In this case, we obtain \cite{Rh3}

\beq
\langle \dot{a}\rangle_{\rm sync} \simeq -\frac{21nC\overline{\epsilon}_\rho e^2}{ma}
\frac{\gamma n}{\gamma^2+n^2} 
\endeq
and
\beq
\langle \dot{e}\rangle_{\rm sync} \simeq -\frac{21nC \overline{\epsilon}_\rho  e} {2ma^2}
\frac{\gamma n}{\gamma^2+n^2}.
 \endeq 

\section{Nonhomogeneous models}

The creep tide model for homogeneous bodies has been successfully applied to an extended range of real problems. However, it was not able to correctly predict the forced libration of Saturn's satellites \cite{Rh3}, The oscillations calculated for these satellites with tidal theories is generally smaller than the results obtained from observations by the space probe Cassini (see \cite{Rh3}). In some cases, as Enceladus, the agreement of the theory and the observations only becomes possible if we assume the existence of a subsurface ocean \cite{Fo19}.

The extension of the creep theory to differentiated bodies is done by assuming that the radial motions of the matter in the neighborhood of the surfaces separating the various layers is dominated by the Newtonian creep of the more viscous of the two layers meeting at that boundary. The layers are assumed to interact. Besides, the friction between two adjacent layers and the mutual gravitational attraction of the layers provide an efficient mechanism for the transfer of angular momentum inside the body \cite{Fo17}\cite{Fo19}. 

Multilayer models involve a large number of free parameters and may only be justified in the case of well observed bodies for which it is possible to constrain a large number of independent parameters. 
     
\section{Conclusions}

We collected in this paper the main equations of the new version of the creep tide theory in the planar case, developed by Folonier et al. \cite{Rh3} and used them to explore some applications to real problems. This new version enhances the similarity and the differences of the creep tide theory and the Darwin theory already evidenced in previous papers where they were jointly presented \cite{LecTi}. 
In the creep theory, the lag and height of the dynamic tide are derived from Physics first principles and the same hydrodynamical law is used for all bodies, no matter if gaseous or stiff, and is enough to show how different can be the dynamic tides in different bodies and dynamical situations. On its turn, Darwin theories need to introduce different ad hoc laws for different bodies and the classical theories adopting a constant time lag (CTL theories) are valid only for gaseous bodies.
Some other crucial differences are emphasized below.

\begin{itemize}
	\item 
	The creep theory allows the dissipated power in the primary to be predicted as a function of the viscosity of the body (via the relaxation factor $\gamma$) and the frequency of the most influential tidal harmonic. It is different from  Darwinian theories, in which the dissipation is proportional to a quality factor $Q$ related to the tide frequency by an ad hoc power law depending on the nature of the body. The dissipation law in the creep theory is universal and may be even applied to predict the dissipation in cases where the relaxation factor $\gamma$ and the frequency of the most influential harmonic ($\nu$ or $n$) have the same order of magnitude, a case to which the used ad hoc power laws of Darwinian theories do not apply.\\      
	
	\item 
	In the creep theory, the synchronization of the rotation of stiff bodies is naturally allowed as consequence of the torques due to the tidal deformation of the body (dynamic tide). At variance with this, in pure Darwinian theories, the rotation is always driven to a supersynchronous state (a.k.a. pseudosynchronous). For instance, in the CTL theories, the relation between the stationary angular velocity of rotation and the companion mean motion is $\Omega=n(1+6e^2)$. Synchronization is only possible if the eccentricity is zero. Otherwise, when $e \ne 0$, in order to have synchronization, it is necessary to introduce ad hoc asymmetries corresponding to a permanent fossil deformation of the body, generally not confirmed by direct observation. Only known exception is Mercury whose rigid structure dominates by 2-3 orders of magnitude any present tidal effects \cite{Gom}.\\
	
	\item 
	The creep theory shows that the lag and the height of the dynamical tide, as well as the angular velocity of rotation of the primary, have short period variations whose consideration cannot be neglected in the neighborhood of synchronization \cite{Rh2, CBLR}.\\
	
\end{itemize}

\section{Appendix A - The relaxation factor and the viscosity}

The relaxation factor ($\gamma$) used in the equation of the Newtonian creep is a function of the uniform viscosity ($\eta$) of the primary. It may be obtained considering the motions near the surface of the primary as a radial fluid flow across the surface and using a simplified version of the Navier-Stokes equation where transverse motions and inertia are neglected  \cite{LecTi}:
\beq
\frac{\partial^2 V_r}{\partial\zeta^2} + \frac{2}{\zeta} \frac {\partial V_r}{\partial\zeta} - \frac{2V_r}{\zeta^2} = \frac{w}{\eta}
\endeq
where $V_r$ is the radial velocity, $\zeta$ is the radius vector of the surface point, $w$ is the local specific weight 
(N.B. $w=-\nabla p$). 
The pressure due to the body gravitation is approximated by the weight of the mass lying above (or 
missing below) the equilibrium surface, that is, $p=-w(\zeta-\rho)$. 

\begin{figure}[h]
	\centerline{\hbox{
			\includegraphics[height=3.5cm,clip=]{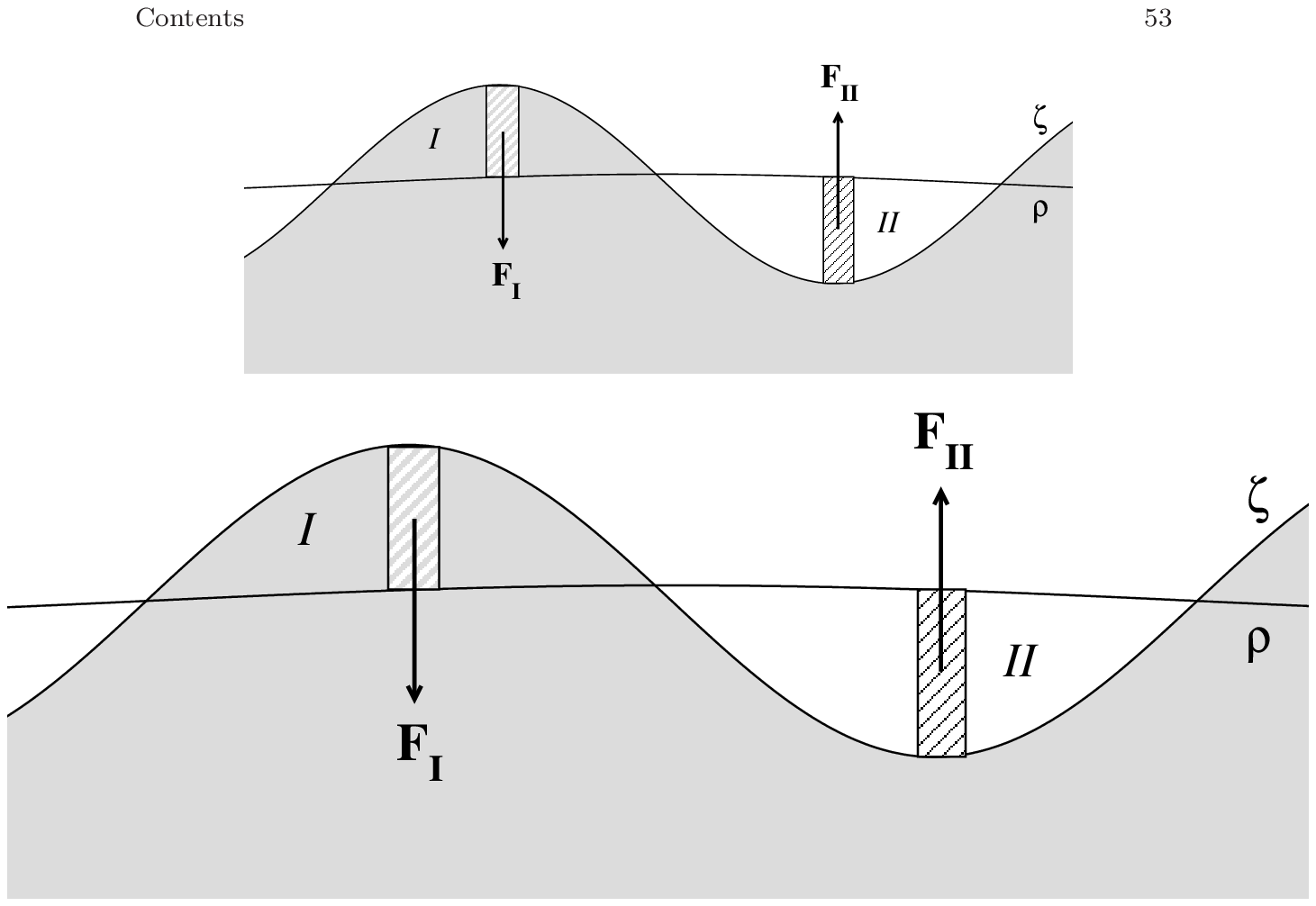}}}
	\caption{The creep model: $\zeta$ is the actual surface of the body at the time $t$ (dynamic tide) and $\rho$ is the surface of the equilibrium ellipsoid at the same time (static tide). Taken from \cite{LecTi}.}
	\label{fig:creep}       
\end{figure}

The solution of this differential equation is: 
\beq
V_r (\zeta) = C_1\zeta  + \frac{C_2}{\zeta^2} - \frac{w}{4\eta}\zeta^2
\endeq
where $C_1$ and $C_2$ are integration constants determined by
the boundary conditions: (i) $V_r(\rho) = 0$ i.e. the velocity vanishes when $\zeta = \rho$ ($\rho$ is the radius vector of the static tide; and (ii) the solution is linear, i.e. $V_r^{''}(\rho) = 0$. Hence
\beq
V_r (\zeta ) = - \frac{wR}{2\eta} (\zeta-\rho) 
\endeq
where $R$ is the body's mean radius. This is the Newtonian creep law with a relaxation factor inversely proportional to the viscosity.

\section{Appendix B - The disturbing force and torque acting on the companion}

When only terms up to degree 2 are taken into account, the potential in a point $\vec{r}\equiv(r,\theta,\varphi)$ due to the gravitational attraction of a homogeneous triaxial ellipsoid whose equatorial plane lies on the fundamental plane of the reference system (coplanar case) is (MacCullagh's formula; see \cite{Beutler}, Sec. 3.3) 
\begin{eqnarray}\label{eq:U}
U(\vec{r}) &=& -\frac{Gm}{r} -\frac{G(B-A)}{2r^5}\bigg(3(\vec{{r}}\cdot\vec{\widehat{r}}_\mathcal{B})^2-r^2\bigg)+\frac{G(C-B)}{2r^3}\bigg(3\cos^2\theta-1\bigg)
\end{eqnarray}
where $G$ is the gravitational constant,  $A,B,C$ are the moments of inertia of the ellipsoid with respect to its principal axes ($A<B<C$), and $\vec{\widehat{r}}_\mathcal{B}$ is the unitary vector oriented to the vertex of the  ellipsoid.

To calculate the disturbing force $\vec{F}$ acting on the companion, we take the negative gradient of the potential of $\tens{m}$ and multiply it by the mass placed in the point. Hence\footnote{The sign in this expression comes from the fact that we are using the conventions of Physics ($\delta U$ is a potential not a force-function). It is important to stress that in agreement with Newton laws, there exists a reaction force $-\vec{F}$ acting on the primary (see \cite{FBM}). This reaction force is generally neglected in studies where one of the masses is much smaller than the other but, in general problems,  its neglect is an error.}
\begin{equation}
\vec{F}=-M\nabla_{\vec{r}} \delta U 
\end{equation}
where $\delta U$ is the disturbing part of $U$. If we take into account that, by hypothesis, the companion lies on the equatorial plane of the primary $\tens{m}$, we may simplify the above equation using $\cos\theta=0$ and $\vec{{r}}\cdot\vec{\widehat{r}}_\mathcal{B}=r\cos(\varphi_B-\varphi)=r\cos\delta$. 
$\varphi_B$ is the angular distance of the considered point to the vertex of the ellipsoid, that is, to the direction of the companion.
  
Hence, if $\vec{F}\equiv(F_1,F_2,F_3)$, we have
\beq\label{eq:F1}
F_1=-M\frac{\partial \delta U}{\partial r}=
 -\frac{3GM(B-A)}{2r^4}(3\cos^2\delta-1) -\frac{3GM(C-B)}{2r^4},
\endeq
\beq\label{eq:F3}   
F_3=-M\frac{1}{r\sin\theta}\frac{\partial \delta U}{\partial \varphi}=
\frac{3GM(B-A)}{2r^4}\sin 2\delta, 
\endeq
and $F_2=0$ (because $\theta=\pi/2$). 
The torque of this force on the companion is $\vec{M}=\vec{r}\times \vec{F}$, or 
\beq\label{eq:torque}
\vec{M} = rF_3\ \vec{k}=\frac{3GM(B-A)}{2r^3}\sin 2\delta \ \vec{k}
\endeq
where $\vec{k}$ is the unit vector along the polar axis of the primary.

The potential, force and torque can be written in terms of the flattenings using the first-order approximations:
\beq
B-A=C\mathcal{E}_\rho, \hspace{1cm}C-B=C(\mathcal{E}_z-\half \mathcal{E}_\rho).
\endeq
Hence, taking into account that $\theta=\pi/2$,
\begin{eqnarray}\label{eq:Ubis}
U(\vec{r}) &=& -\frac{Gm}{r} -\frac{3GC\mathcal{E}_\rho}{4r^3}\cos 2\delta -\frac{GC\mathcal{E}_z}{2r^3},
\end{eqnarray}
\beq\label{eq:F1bis}
F_1=-\frac{9GMC\mathcal{E}_\rho}{4r^4}\cos 2\delta -\frac{3GMC\mathcal{E}_z}{2r^4},
\endeq
\beq
F_2=0, 
\endeq
\beq\label{eq:F3bis}   
F_3=\frac{3GMC\mathcal{E}_\rho}{2r^4}\sin 2\delta, 
\endeq
and 
\beq\label{eq:torquebis}
\vec{M} =\frac{3GMC\mathcal{E}_\rho}{2r^3}\sin 2\delta \ \vec{k}.
\endeq

\section{Appendix C - The Earth conundrum}

One result of the creep tide theory is the near $45^\circ$ lag of the tidal bulges of free rotating (i.e. non synchronized) stiff bodies. This result is surprising  because since Darwin's early studies of the Earth rotation, the lag of the tidal bulge is assumed to be small (as it indeed is in the case of gaseous bodies). This is not the only divergence. In the creep tide theory, $E_\rho$ is also variable and, in the case of free rotating stiff bodies, it is very small. It is not a finite quantity as in Darwinian theories.

However, notwithstanding these two diametrically opposed results, both theories predict similar rotational and orbital evolutions. This is easy to understand. The rotational evolution, for instance, is ruled by Eq. (\ref{eq:Folo4}) in both theories, which  expresses the angular acceleration of an homogeneous ellipsoid under the gravitational pull of an external body, independently of the physics leading to the formation of the ellipsoid. It shows that the acceleration is proportional to $E_\rho\sin 2\delta$. In the creep tide theory $\sin 2\delta \simeq 1$ and $E_\rho$ is a small quantity whose actual value depends on the relaxation factor $\gamma$, that is, on the viscosity. In Darwinian theories, $E_\rho$ is given and $\delta$ is an ad-hoc quantity. In classical versions of the theory, $E_\rho \sim 1 $, but in more realistic ones, $E_\rho$ is modified by an attenuation factor $\cal A$ to take into account the density distribution in the interior of the body and the incomplete deformation of the Earth under the moving tidal stress. 
If the free parameters are chosen such that 
\begdi
\delta_{\rm Darwin} = \frac{\gamma}{2{\cal A}\sqrt{\gamma^2+\nu^2}},
\enddi
the two theories will lead to the same rotational evolution. In addition, if the eccentricity is small, the orbital evolutions given by the two theories will be close one from another. However, evolution time scales are large and we do not have  observational access to the actual evolution of the related quantities with enough precision to select one theory and discard the other.

The only way to decide on one of these theories is through the experimental determination of $E_\rho$ or $\delta$. It happens that the Earth has oceans that strongly affect tidal responses and it is very difficult, if not impossible, to disentangle the two effects to have a determination corresponding to the solid Earth alone. 

The following facts enhance the contradictions: 
\begin{itemize}
	\item 
	The viscosity of the mantle is very close to $10^{21}$ Pa s \cite{Cath}. As a consequence, if the fluidity of the core is neglected, $\gamma \sim 1.7 \times 10^{-10}\ {\rm s}^{-1}$ and the resulting $E_\rho$ is too small. The resulting bulge height is less than millimetric.   
	\item 
	The tidal wave ${\rm O}_1$ (diurnal lunar) obeys the static theory and is not much disturbed by oceanic indirect effects or by atmospheric perturbations as its period is 25h 49m \cite{Melc}. The results from measurements done with horizontal pendulums and gravimeters gives for the tidal effective Love numbers,
	$k = 0.317 \pm 0.011$ and
	$h = 0.638 \pm 0.017$.
	If we use the standard formula \cite{Fo15}
	\beq
	\frac{a-b}{R} = \frac{2}{5} h \epsilon_\rho,
	\endeq
	we obtain $a-b=16$ cm.
	\item 
	VLBI observations confirm the value $h=0.600 \pm 0.001$ of this Love number, for the tidal wave ${\rm M}_2$ (semidiurnal lunar). \cite{Haas}
	\item 
	In order to overcome the influence of the ocean in the determination of the lag of the solid Earth, Ray et al. \cite{Ray96} have combined Topex-Poseidon altimeter measurements of ocean tide with the laser tracking of satellites orbit perturbations to determine $k \sin\delta$. Using the above values of $k$, they obtained $\delta = 0.016 \pm 0.009$ degrees. This result was later improved \cite{Ray} to  $\delta = 0.020 \pm 0.005$ degrees.
	
\end{itemize}
These contradictions show the impossibility of treating one problem so complex as the solid Earth tides with the tidal theories built to deal with homogeneous bodies. In the case of the creep tide theory, this difficulty was overcome in \cite{Rh1} by the ad hoc addition of an elastic term to the results obtained with the theory. making its results akin to those obtained with Maxwell viscoelastic models \cite{SFM-AA}. An alternative solution may also exist using a layered model and taking into account the low viscosity of the LVZ (low-velocity zone) in the upper asthenosphere, just beneath the lithosphere \cite{Dogl}.


\begin{acknowledgement}
	\textit{Acknowledgement}. We thank M. Efroimsky for suggestions concerning some statements on Darwin's theory. This investigation is funded by the National Research Council, CNPq, grant 302742/2015-8 and by FAPESP, grants 2016/20189-9 and 2017/25224-0. This investigation is part of the thematic project FAPESP 2016/13750-6. 
	Authors contribution statement: The authors are working together on the subjects covered by this paper since many years. 
\end{acknowledgement}



\end{document}